\documentclass[%
 aip,
 amsmath,Amssymb,
 reprint,%
jap,groupedaddress]{revtex4-1}

\usepackage{graphicx}
\usepackage{dcolumn}
\usepackage{bm}
\usepackage{slashed}

\usepackage[utf8]{inputenc}
\usepackage[T1]{fontenc}
\usepackage{mathptmx}
\usepackage{etoolbox}

\makeatletter
\def\@email#1#2{%
 \endgroup
 \patchcmd{\titleblock@produce}
  {\frontmatter@RRAPformat}
  {\frontmatter@RRAPformat{\produce@RRAP{*#1\href{mailto:#2}{#2}}}\frontmatter@RRAPformat}
  {}{}
}%
\makeatother
\begin{document}

\title{ Boundary Ferromagnetism in Zigzag Edged Graphene}
\author{Gordon W. Semenoff}
 \affiliation{Department of Physics and Astronomy, University of British Columbia\\ 6224 Agricultural Road, Vancouver, British Columbia, Canada V6T 1Z1 }


\begin{abstract}
The flat band of edge states which occur in the simple tight-binding lattice model of graphene with a zig-zag edge have long been conjectured to take up a ferromagnetic configuration.  In this work we demonstrate that, for a large class of interaction Hamiltonians which can be added to the tight-binding model, and  at the first order in perturbation theory,  the degeneracy of the edge states is resolved in such a way that the ground state is in the maximal, spin $j=N/2$ representation of the spin symmetry where $N$ is the number of edge states. 
 \end{abstract}

\maketitle

\section{Introduction and Summary}

One of the fascinating features of graphene with a zig-zag edge is the  appearance of  edge states for the graphene electrons.  The number of these states is semi-macroscopic --  it is  roughly equal to the number of atomic sites at the edge.   In the tight binding model of graphene they constitute a perfectly flat energy band sitting at the charge neutral point, at the same energy as the apexes of the Dirac cones that give graphene its relativistic electron spectrum\cite{gs}.  

The edge state solutions of the tight binding model were noticed long ago \cite{zig,zig1},  and their interesting properties and important potential applications have inspired literally thousands of research papers since then.  The existence of the edge states themselves has been confirmed by experiment \cite{exp_12,exp_13,exp_14}.  This confirmation lends some credence to the simple tight binding model which predicts them.   

 One of the properties of the edge states that has been conjectured from the beginning is ferromagnetism, the alignment of the spins of the electrons which populate the states.  This idea was put forward early on 
using mean field theory with a simple Hubbard interaction\cite{zig}.  It has been 
supported by various approximate computations using mean field theory\cite{edgeferro_mft}, density functional\cite{edgeferro_dft} and numerical techniques\cite{edgeferro_num}.  Given the  important potential  applications of the edge magnetism in spintronics \cite{edgeferro_dft,spin1,spin2} and the fact that it has not been seen by experiments yet, it is important to gain a better understanding of this phenomenon.  

An important piece of the puzzle comes from the application of Lieb's theorem\cite{Lieb}  to the bipartite graphene lattice, where the Hamiltonian is taken to be that of the tight binding model with a repulsive Hubbard interaction added \cite{l1,l2,l3,Affleck1}.  This theorem states that, at half filling, the ground state of the Hubbard model on a bipartite lattice  is $2j+1$-fold degenerate and it carries an irreducible representation of the $su(2)$ spin algebra with $j=\frac{1}{2}||A|-|B||$, where $|A|$ and $|B|$ are the numbers of $A$ and $B$ sites.  There is no further degeneracy of the ground state. This is compatible with the exact solutions of the tight-binding model of graphene which is the Hubbard model with the Hubbard interaction switched off. In that limit, for zigzag and bearded edges, an entire flat band of electronic states appears at the Fermi level. Then any half-filling of that flat band has the same energy as any other half-filling. The flat band generally has $2j$ single electron states, $4j$ with spin degeneracy included,  and partial fillings of the flat band have $2^{4j}$ multi-electron states, $(2^{4j})!/(2^{4j}-1)^2$ of which are half-filled. 

An interesting implication of Lieb's theorem is the expectation that, if we added even an infinitesimally weak repulsive Hubbard interaction to the standard tight-binding Hamiltonian of graphene, 
this enormous degeneracy of  $[(2^{4j})!]/[(2^{4j}-1)!]^2$ multi-electron states would be lifted, leaving an essentially unique $2j+1$-fold degenerate ground state.  

The other interesting implication is that those $2j+1$ remaining states transform under the spin $j$ irreducible representation of the $su(2)$ spin algebra.  For a macroscopic sample of graphene with a zigzag edge, $j$ is large.  It is approximately $\frac{2}{3}$ times the number of atoms on the edge.  In any state in an irreducible representation of the $su(2)$ algebra the spin is polarized in some direction.  What is more, when the representation is large, $j\to\infty$, this spin should behave semi-classically, leading to spontaneously broken spin symmetry and a ferromagnetic state.   
 
 There remains the question as to whether the edge magnetization is compatible with other interactions which are longer ranged than the zero-range Hubbard interaction.  An example is the Coulomb force, which is perhaps
 the most important interaction in any realistic modelling of graphene and for which Lieb's theorem is not applicable. This question was addressed by Shi and Affleck\cite{Affleck2} who argued that the Coulomb interaction, when projected onto the edge states, still splits the energy levels of the half-filled flat band so that the ground state is a ferromagnet. Their argument was limited to the first order of perturbation theory and it ignored interactions of the edge states with the bulk states entirely. They concluded that the same resolution of the degeneracy occurs for a weak Coulomb interaction as would have occurred for a weak Hubbard interaction and that the resulting $2j+1$ states are spin polarized.   This scenario has since been supported by quantum Monte Carlo simulations of such a system\cite{qmc} where they find that the magnetic order indeed appears tand it persists with no sign of a phase transition as the long range interaction strength is increased through some range up to finite values.  Of course the strong interaction limit for a the Coulomb force should eventually result in an antiferromagnetic Mott insulator\cite{sem} which one would expect is separated from the Dirac semi-metal by a phase transition. 
 
 In this paper, we will exploit some techniques which were developed in investigations of quantum Hall ferromagnetism\cite{sz} and boundary conformal field theory\cite{bs} to revisit the problem of the splitting of the flat band degeneracy due to a weak repulsive interaction.   We shall show that, for a large class of interaction potentials including the repulsive Hubbard and Coulomb interaction, to the first order in degenerate perturbation theory, the degeneracy of the flat edge band is indeed resolved so that the ground state is a single  spin $j$ representation of the $su(2)$ spin algebra, where $2j$ is the number of single-electron edge states of the tight-binding Hamiltonian.  We include the direct and exchange interactions of the edge states with the electrons and holes in the bulk.  The latter interactions are generally not small, and especially when they are long ranged, they cannot be legitimately neglected.   Our work has significant overlap with that of Shi and Affleck\cite{Affleck2} and where we overlap we agree with them.  What we add to the subject is the complete analysis of interactions between the edge and the bulk degrees of freedom.  Indeed they turn out to be important.  We use sum rules and particle-hole symmetry to refine them so that the end result  leaves what is basically the same problem as the projection of the interaction onto edge states with the appropriate guess for the off-set of the edge state charge density, which might be guessed by requiring charge neutrality. We also refine the proof, originally given by Affleck and Karimi\cite{Affleck1}, that the lowest energy states are spin polarized. This will make use of an emergent $su(2j)$ Lie algebra that we will show the ground states must carry a trivial representation of. 
   
 We will limit our consideration to interaction Hamiltonians which contain a spin-independent two-body interaction $\mathcal V(X,Y)$ so that, when written in terms of the  creation and annihilation operators of electrons, $\psi_\sigma^\dagger(X)$ and $\psi_\sigma(X)$ respectively, it has the form
\begin{align}
H_{\rm int}= \frac{1}{2}\sum_{XY}~\mathcal V(X,Y)~\rho(X)  \rho(Y) 
\label{hint} \\
\rho(X)~=~\sum_{\sigma=1}^2\psi_\sigma^\dagger(X)\psi_\sigma(X)-1 
\label{rho}
\end{align}
Here, $X$ and $Y$ are the positions of lattice sites and $\sigma$ labels the two spin states of the electron. If we interpret the density $\rho(X)$ as being proportional to the electric charge density at lattice site $X$, the ``$-1$''  is due to the charge of the ion residing at each lattice site.  Since the electron has two spin states, the charge neutral state has an average electron density of one electron per site.   In other cases it should be regarded as a chemical potential that is tuned to a convenient value.  We will always assume that the electronic states are half-filled.  

The two-body interaction potential $\mathcal V(X,Y)$ will be assumed to be  symmetric and positive.  It need not be translation invariant.  
Positivity of the potential is defined by the spectral problem
\begin{align}
\sum_Y \mathcal V(X,Y)~\varphi_v(Y)~=~v~\varphi_v(X)
\label{pot1}
\end{align}
The kernel $ \mathcal V(X,Y)$ is positive if all of its eigenvalues $v$ are positive. The potential has an eigenfunction decomposition
\begin{align}
\mathcal V(X,Y)=\sum_v~v~ \varphi_v(X)\varphi^*_v(Y) 
\label{pot2}
\end{align}

Examples of such an interaction include many which are commonly used to model the interactions of electrons
in graphene and other Dirac materials.  Important ones are the repulsive Hubbard interaction
\begin{align}
\mathcal V_{\rm Hubbard}(X,Y)= U_0 \delta(X,Y)
\end{align}
with $U_0 >0$ and the Coulomb interaction
\begin{align}
\mathcal V_{\rm Coulomb}(X,Y)=\biggl\{ \begin{matrix} U_0 & X=Y \cr  \frac{e^2}{4\pi\epsilon|X-Y|} & X\neq Y\cr \end{matrix}
\end{align}
 A further, less easy to quantify property of the interaction that we need is the accuracy of the leading order of perturbation theory. 
Quantitatively, this means that all matrix elements of the interaction Hamiltonian in the relevant multi-electron states should be smaller than
one the energy scale of the tight-binding Hamiltonian (the parameter $t$ in the tight-binding Hamiltonian (\ref{ham}) below). This is indeed the case
for the Hubbard and Coulomb interactions listed above, but it is not so for every positive potential, for example, one whose strength grows with distance
would eventually always be a strong interaction of the system is large enough.

\begin{figure}[h!]
\centerline{\includegraphics[scale=.8]{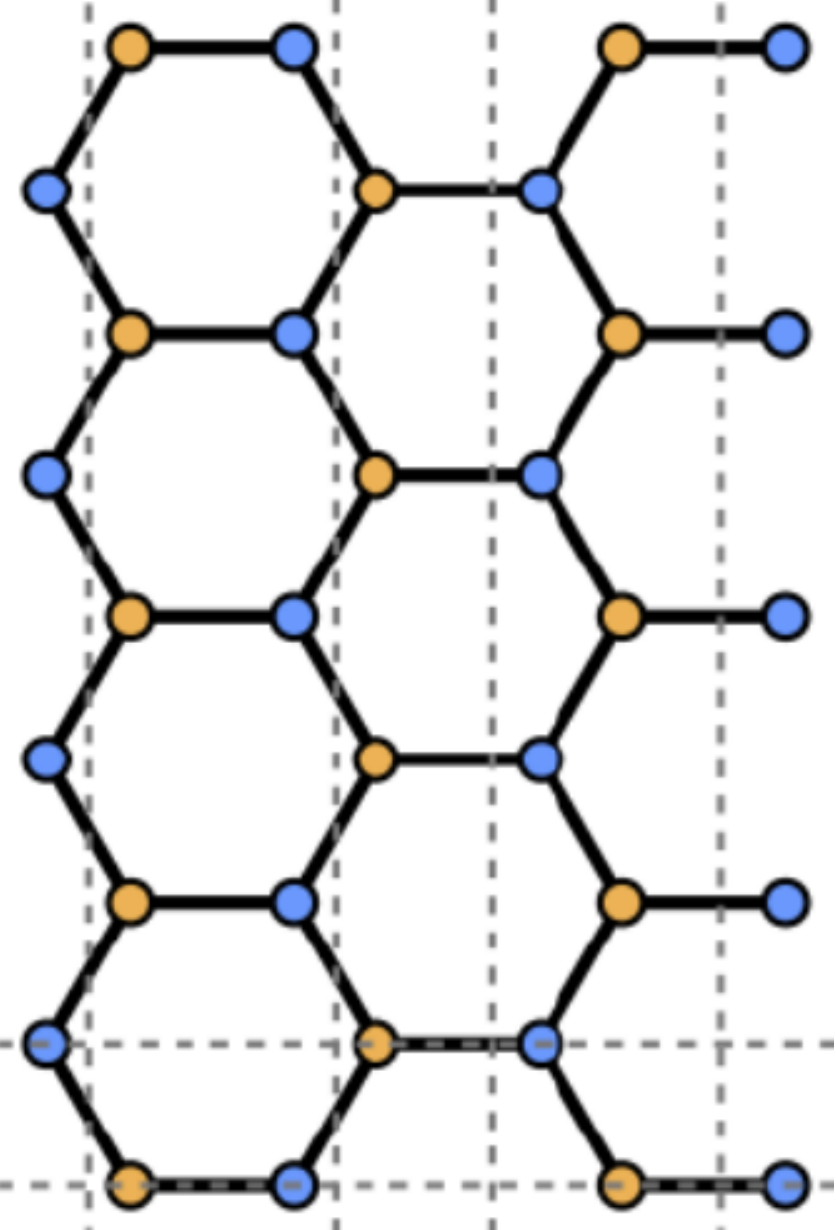}}
\caption{\small The hexagonal graphene lattice is depicted.  The blue dots are the $A$ sub-lattice and the red dots are the $B$ sub-lattice. 
The   zigzag is on the left-hand-side and the lattice is assumed to have indefinite extension to the right.  The up-down directions will either be assumed to be infinite or to 
have a periodic identification by $L$ units where $L$ is an integer.  The semi-infinite sheet can be recovered by taking $L\to\infty$. }
\label{fig}
\end{figure}
We will also confine our attention to a semi-infinite sheet of graphene having a single zigzag edge, the configuration which is depicted in figure \ref{fig}.   It should be easy to generalize what we do to 
a system with bearded edges or a nano-ribbon with one zigzag and one bearded edge.  It would also apply to a nano-ribbon with two zigzag edges which is sufficiently wide
that the interaction of the edges can be ignored, in which case the properties of each edge would be as if it were the edge of a semi-infinite sheet.
We will use a periodic identification of the system in the direction parallel to the edge.  This helps us in that the number of edge states is then finite and the space of quantum states that we must study will live in a finite dimensional complex vector space.  The semi-infinite sheet can be recovered by simply taking the period of the identification to infinity.   On the other hand, our results do apply to the periodically identified sheet which would to some approximation
 describe a semi-infinite nano-tube with a zig-zag edge at its cap.

 The remainder of this paper contains a detailed exposition of the results that we have outlined above.  In section \ref{tight-binding} we will review solutions of the tight-binding model in a semi-infinite graphene sheet with a zigzag boundary.
 The content of this section is well-known and can be found in many places in the literature\cite{review}.  We review it here simply to establish our notation and for the convenience of the reader. Section III contains our main results which we have outlined above.    Section IV has a summary and further discussion.  Some of the  technical  details are summarized in the Appendices.

 \section{ Tight binding model}
 \label{tight-binding}
 The tight binding model of graphene has electrons occupying sites of the bipartite honeycomb lattice, a caricature of which is depicted in figure \ref{fig}. A review of its structure and details of our notation and conventions for lattice and dual lattice vectors are summarized in Appendix \ref{lattice}.  The hexagonal lattice is made up of two triangular sub-lattices which we  call the $A$ and $B$ sub-lattices. This lattice is populated by electrons and the charge neutral system has one electron per site. 
 The tight binding model of graphene has the Hamiltonian
\begin{align}
 H_0=\sum_{A}\sum_{\sigma=1}^2\sum_{i=1}^3 \biggl[ t^* \psi^\dagger_\sigma(A+\hat \delta_i)\psi_\sigma (A)
  +
 t \psi_\sigma^\dagger(A) \psi_\sigma(A+\hat \delta_i)\biggr]
 \label{ham}
 \end{align}
where $A$ denotes sites on the triangular A sub-lattice and $\hat\delta_i$ are displacements between an $A$ site and its nearest neighbours which are on the $B$ sub-lattice.    Each link of the lattice occurs twice in the Hamiltonian, once with amplitude $t^*$ for hopping from an $A$ site to a neighbouring $B$ site and one with amplitude $t$ for the inverse process.  We could without loss of generality choose $t$ to be real and positive, and we will do so in the following.  This is a manifestation of time reversal symmetry of all of the models that we will consider. 

The operators $\psi^\dagger_\sigma(X)$ and $\psi_\sigma(X)$ create and annihilate an electron in spin state $\sigma$ at site $X$. They have the anti-commutator algebra
  \begin{align}
 & \biggl\{ \psi_\sigma(X),\psi^\dagger_\tau(Y)\biggr\}=\delta_{\sigma\tau}\delta(X,Y) ,~\nonumber \\ &   \biggl\{ \psi_\sigma(X),\psi_\tau(Y)\biggr\}=0,~
   \biggl\{ \psi^\dagger_\sigma(X),\psi^\dagger_\tau(Y)\biggr\}=0
   \label{commutator}
   \end{align}
  
We will consider this system on a half-space where, in figure \ref{fig}, the zigzag edge is on the left-hand-side and the edge sites are located entirely on the $A$ sub-lattice. The edge is taken into account by imposing a boundary condition for the lattice field,  
\begin{align}
\psi_\sigma(B_1=0,B_2)=0
\label{bc}
\end{align} 
The tight-binding model Hamiltonian (\ref{ham}), and all of the interactions that we will consider,  have   $su(2)$ spin symmetry in that the Hamiltonians
commute with the generators of the $su(2)$ Lie algebra which can be taken to be
\begin{align}
&J^1=\frac{1}{2}\sum_X[\psi^\dagger_1(X)\psi_2(X)+\psi^\dagger_2(X)\psi_1(X) ]\\
&J^2=- \frac{i}{2}\sum_X[\psi^\dagger_1(X)\psi_2(X)-\psi^\dagger_2(X)\psi_1(X)]\\
&J^3=\frac{1}{2}\sum_X[\psi^\dagger_1(X)\psi_1(X)-\psi^\dagger_2(X)\psi_2(X)]\\
&\left[J^a,J^b\right] = i\epsilon^{abc}J^c\\
&\left[ J^a,H_0\right]=0~,~\left[ J^a,H_{\rm int}\right]=0
\end{align}
where $H_{\rm int}$ is the operator in equation (\ref{hint}). This implies that quantum states can be organized into irreducible representations of $su(2)$.  We will find this fact useful. 
The tight-binding model Hamiltonian (\ref{ham}), and all of the interactions that we will consider,  also have  a $U(1)$  symmetry corresponding to the conservation
of electric charge. 
\begin{align}
&Q= \sum_X\left[ \psi^\dagger_1(X)\psi_1(X)+ \psi^\dagger_2(X)\psi_2(X)\right]  \\
&\left[ Q,H_0\right]=0~,~\left[ Q,H_{\rm int}\right]=0
\end{align}

 To solve the tight-binding model with Hamiltonian (\ref{ham}), we note that the Heisenberg equations of motion (for Heisenberg picture fields) that can be derived from the Hamiltonian (\ref{ham}) are
 \begin{align}
 i\hbar\frac{d}{d\tau}  \psi_\sigma(A,\tau) =\left[ \psi_\sigma(A,\tau),H\right]
 =  t\sum_i  \psi_\sigma(A+\delta_i 
 ,\tau) \label{se1}\\
  i\hbar\frac{d}{d\tau}  \psi_\sigma(B,\tau)  =\left[ \psi_\sigma(B,\tau),H\right]
 =t\sum_i 
  \psi_\sigma(B-\hat\delta_i ,\tau) \label{se2}
 \end{align}
 
 It is easy to find solutions of these equations.  We separate the time by making the ansatz
  \begin{align}
 \psi (A,\tau)=\phi(A)e^{ -i\omega \tau},~
 \psi (B,\tau)=\phi (B)e^{-i\omega \tau}
 \end{align}
 and find wave-functions by taking superpositions of plane waves which satisfy the satisfy the resulting difference equation with the boundary condition (\ref{bc}).
 We find  a positive frequency band of solutions, with frequency and wave-functions given by
\begin{align}
 &\omega(k)~=~\frac{t}{\hbar} |S(k)|\label{1/2+0}\\
&\phi^{(+)}(k;B)=\frac{2}{\sqrt{2\Omega} }e^{ik_2B_2} \sin(k_1B_1) \label{1/2+1}\\
&\phi^{(+)}(k;A) = \frac{e^{ik_2A_2}}{\sqrt{2\Omega} i} \biggl[ e^{ik_1A_1}\frac{S^*(k)}{|S(k)|} - e^{-ik_1A_1}\frac{S(k)}{|S(k)|}  \biggr]
\label{1/2+}\end{align}
and a negative frequency band with frequency and solutions 
 \begin{align}
 &\omega(k)~=~-~\frac{t}{\hbar} |S(k)| \label{1/2-0}\\
&\phi^{(-)}(k;B)= \frac{2}{\sqrt{2\Omega}}e^{ik_2B_2} \sin(k_1B_1) \label{1/2-}\\
&\phi^{(-)}(k;A) = ~-~
 \frac{ e^{ik_2A_2} }{\sqrt{2\Omega} i}\biggl[ e^{ik_1A_1}\frac{S^*(k)}{|S(k)|} - e^{-ik_1A_1}\frac{S(k)}{|S(k)|}  \biggr]
\label{1/2-}\end{align}
where $\Omega$ is the volume of the Brillouin zone of one of the triangular sub-lattices (see Appendix \ref{lattice}), $k\in\Omega^+$ takes values in the half of the Brillouin zone which has $k_1>0$ and 
 \begin{align}
 S(k)=\sum_i e^{i\vec k\cdot\hat\delta_i}=e^{-ik_1}+e^{i\frac{1}{2}k_1}2\cos\left(\frac{\sqrt{3}}{2}k_2\right) \label{s}
 \end{align}
 We shall call the positive and negative frequency solutions (\ref{1/2+0})-(\ref{1/2+}) and (\ref{1/2-0})-(\ref{1/2-}) the ``bulk states''. For future reference we note that the positive and negative energy bulk states are very similar.  They differ only by a flip of the sign of the part that is supported in the $A$ sub-lattice.  This is a manifestation of particle-hole symmetry.

In addition to the positive and negative energy bands of bulk states, there is a flat band of  zero frequency solutions which we shall call edge states, 
\begin{align}
 &\omega~=~0\\
&\phi^{(0)}(B)~=~0 \\
&\phi^{(0)}(k_2;A) =\frac{ \sqrt{1-4\cos^2(   \frac{\sqrt{3}k_2}{2} ) }}{\sqrt{2\pi/\sqrt{3}}}   \biggl[-2\cos(   \frac{\sqrt{3}k_2}{2})   \biggr]^{\frac{2}{3}[A_1-1]}   e^{ik_2A_2} 
\label{e}
\end{align}
 This solution  has support only on the $A$ sub-lattice and the wave-number $k_2$ takes values such that   
 \begin{align}\label{k20}
 -1<2\cos(   \frac{\sqrt{3}k_2}{2})<1
 \end{align}
 The values of $k_2$ such that $e^{ik_2A_2}$ are linearly independent are 
 \begin{align}\label{k21}
 0\leq \frac{\sqrt{3}}{2}k_2<\pi 
 \end{align}
  and
 then the   condition (\ref{k20}) tells us that  $k_2$ for edge states must be in the sub-interval
\begin{align}\label{k22}
 \frac{\pi}{3}<\frac{\sqrt{3}k_2}{2}<\frac{2\pi}{3}
 \end{align}
  The wave-functions of the edge states are normalized so that 
 \begin{align}
\sum_{A^+}\phi^{(0)}(k_2;A) \phi^{(0)*}(k_2';A) = \delta(k_2-k_2')
\end{align}
where $A^+$ denotes the $A$ sites with $A_1\geq 0$.  The Dirac delta function  in the above expression is periodic with period equal to  the dual vector to the one dimensional lattice on the edge, that is $\delta(k)=\delta(k+2\pi/\sqrt{3})$.   

We will also generally consider a periodic identification of the lattice in the $A_2,B_2$ directions
\begin{align}\label{identification}
A\sim A+L(0,\sqrt{3})~,~B\sim B+L(0,\sqrt{3})
\end{align}
where $L$ is a positive integer.  
What this modification does for us is to make $k_2$ discrete and it takes on a finite number of values in
the interval in equation (\ref{k21}) or, in the case of edge states, in the interval in equation (\ref{k22}).  

This discreteness is found by  requiring periodicity of $e^{ik_2A_2}$ and $e^{ik_2B_2}$ 
under the identification (\ref{identification}). The result is
\begin{align}
&\frac{\sqrt{3}}{2}k_2= \frac{\pi}{L}\ell~,~\ell=0,2,...,L-1
\label{k23}
\end{align}
These are the allowed discrete values of $k_2$ for the bulk states. 
 In the following we will usually denote the 
sum over the allowed values of $k_2$ listed in equation (\ref{k23}) which are arguments of  wave-functions or  creation and annihilation operators by an integral but in all cases, this is easily converted to the appropriate sum of $\ell$ over its range. We will make interchangeable use the notations
$$
\int dk_2~~ \leftrightarrow ~~\frac{2\pi}{\sqrt{3}L}\sum_{\ell=0}^{L-1}~~({\rm sum~of~bulk~states})
$$

If $k_2$ appears in an edge wave-function or edge state creation or annihilation operator,  $\ell$ must be taken in the 
smaller domain
\begin{align}
&\frac{\sqrt{3}k_2}{2}= \frac{\pi}{L}\ell~,~\ell \in\mathcal Z~\bigwedge~\ell \in (L/3,2L/3)
\label{k24}
\end{align}
Here $\ell$ must occur in the open interval since if it were equal to one of the endpoints (only possible when $L$ is a  multiple of $3$) 
the edge state wave-function would not be normalizable.   Also, if $L$ were even, $\ell=L/2$ is not allowed since the wave-function vanishes there. 
These are interesting issues which are not in the main line of our arguments in this paper.  We could avoid them by simply insisting that $L$ is a positive integer which is not a multiple of 2 or 3. We will denote the number of values of $\ell$ allowed by equation (\ref{k24}) and $\ell\neq L/2$ by the integer $N$. We will often use the following two notations for a sum over these allowed values of $k_2$ interchangeably,
$$
\int dk_2~~ \leftrightarrow ~~\frac{2\pi}{3\sqrt{3}N}\sum_{\ell}~~({\rm sum~of~edge~states})
$$
Our purpose for this periodic identification is to make the number of edge states finite.  This helps to properly define the space of quantum states and we could later take the limit where $L$ and $N$ go to infinity in order to recover the latticized half-plane. On the other hand, the system with $L$ and $N$ finite also has interesting applications where it describes a semi-infinite nanotube with a zigzag edge. 

It is easy to confirm that the wave-functions obey the completeness relations
\begin{align}
&\int_{\Omega^+}dk_1dk_2\biggl[  \phi^{(+)}(k;B)\phi^{(+)*}(k,B')+   \phi^{(-)}(k;B)\phi^{(-)*}(k,B')\biggr]
\nonumber \\
&
=\delta(B,B')
\label{c1}
\\
&\int_{\Omega^+}dk_1dk_2\biggl[  \phi^{(+)}(k;A)\phi^{(+)*}(k,A')+   \phi^{(-)}(k;A)\phi^{(-)*}(k,A')\biggr]
\nonumber \\
&+
\int  dk_2  \phi^{(0)}(k;A)\phi^{(0)*}(k,A') 
=\delta(A,A')
\label{c2}\\
&\int_{\Omega^+}dk_1dk_2\biggl[  \phi^{(+)}(k;A)\phi^{(+)*}(k,B)+   \phi^{(-)}(k;A)\phi^{(-)*}(k,B)\biggr]
\nonumber \\ &=0
\label{c3}
\end{align}
 The only one of these which is nontrivial is equation (\ref{c2}) and it is derived explicitly in Appendix \ref{completeness}.
These equations imply that we have all of the solutions.  

Given that we have solutions of the Schr\"odinger equation that follows from the tight-binding Hamiltonian, we can write the lattice fields in a mode expansion, 
\begin{align}
&\psi_\sigma (X,\tau)=\int dk_2  \phi^{(0)}(k_2,X)c_\sigma(k_2) + \nonumber \\ &
\int_{\Omega^+} dk_1dk_2 \left[
\phi^{(+)}(k;X)e^{-i\omega(k)\tau}a_\sigma(k)
+\phi^{(-)*}(k;X)e^{i\omega(k)\tau}b^\dagger_\sigma(k)\right] \label{resolution} \\ &
=\psi_\sigma^{(0)}(X.\tau)+\tilde\psi_\sigma(X,\tau) \label{decomposition} \\ &
\psi_\sigma^{(0)} (X,\tau)=\int dk_2  \phi^{(0)}(k_2,X)c_\sigma(k_2) \nonumber \\ & 
\tilde \psi_\sigma (X,\tau)=\nonumber \\ &
\int_{\Omega^+} dk_1dk_2 \left[
\phi^{(+)}(k;X)e^{-i\omega(k)\tau}a_\sigma(k)
+\phi^{(-)*}(k;X)e^{i\omega(k)\tau}b^\dagger_\sigma(k)\right] \nonumber
\end{align}
where we separate the field containing the edge degrees of freedom, which we denote as $\psi_\sigma^{(0)}(X,\tau)$ and the part containing the bulk degrees of freedom, which we denote by $\tilde\psi(X,\tau)$. The creation and annihilation operators obey the algebra whose non-vanishing anti-commutators are
\begin{align}
&\left\{ a_\sigma(k),a_\rho^\dagger(\ell)\right\}=\delta_{\sigma,\rho}\delta(k,\ell) \label{coma}\\
&\left\{ b_\sigma(k),b_\rho^\dagger(\ell)\right\}=\delta_{\sigma,\rho}\delta(k,\ell) \label{comb}\\
&\left\{ c_\sigma(k_2),c_\rho^\dagger(k_2')\right\}=\delta_{\sigma,\rho}\delta(k_2,k_2')\label{comc}
\end{align}
We have presented the time-dependent Heisenberg fields in equation (\ref{resolution}). The creation and annihilation operator algebra in (\ref{coma})-(\ref{comc}) gives the field in (\ref{resolution}) the equal-time commutation relation that is appropriate to such Heisenberg fields.  However, in the following, we shall only need the Scrh\"odinger picture operators which we get from the Heisenberg operators by simply setting the time $\tau=0$. Then the equal time commutation relation reduces to the one quoted in equation (\ref{commutator}) and equation (\ref{resolution}) with $\tau$ set to zero, together with equations (\ref{coma})-(\ref{comc}) and completeness of the wave-functions
(\ref{c1})-(\ref{c3}) are sufficient to produce (\ref{commutator}).

Plugging the solution (\ref{decomposition}) into the tight-binding Hamiltonian gives the expression
\begin{align}
&H_0= \int_{\Omega^+}dk_1dk_2\sum_{\sigma}~t|S(k)| ~\left[ a_\sigma^\dagger(k)a_\sigma(k)+
 b_\sigma^\dagger(k)b_\sigma(k) \right] \nonumber \\ &~~~~~~~~~~~ +E_0 \label{ham1} \\ 
 &E_0=- 2\sum_X \int_{\Omega^+}dk_1dk_2~t|S(k)|  \phi^{(-)}(k;X)\phi^{(-)*}(k;X)
 \label{E0}
 \end{align}
 This operator commutes with the operators $c_\sigma(k_2)$ and $c_\sigma^\dagger(k_2)$. 

We will consider the space of quantum states as a direct product of two spaces, one  of the members of the product carrying a representation of 
the anti-commutator algebra of the operators $a_\sigma(k),a_\rho^\dagger(k), b_\sigma(k),b_\sigma^\dagger(k)$ and other carrying a representation 
of the anti-commutator algebra of the operators $c_\sigma(k_2),c_\sigma^\dagger(k_2)$. 
The algebra of the operators 
$a_\sigma(k),a_\sigma^\dagger(k), b_\sigma(k),b_\sigma^\dagger(k)$  will have the standard Fock space representation which begins with 
 the cyclic vector $|0>$ with  the property $<0|0>=1$ and 
\begin{align}
 a_\sigma(k)|0>=0,~b_\sigma(k)|0>=0~\forall k,\sigma
\end{align}
A basis for the Fock space can be taken as $|0>$ and the vectors that are made by creation operators acting on $|0>$,
$$
\left\{|0>,a^\dagger_\sigma(k)|0>,b^\dagger_\sigma(k)|0>,a^\dagger_{\sigma_1}(k_1)a^\dagger_{\sigma_2}(k_2)|0>, \ldots\right\}
$$
These correspond to the vacuum and electron and hole states of the bulk degrees of freedom. 

The representation of the algebra of the  operators $c_\sigma(k_2),c_\sigma^\dagger(k_2)$ will also be a Fock space, however, we 
 will put off a discussion of the representation of the $c$'s until we identify the states with the lowest energies. 
We note that, when we impose the periodicity conditions (\ref{identification}), $k_2$ takes on $N$ discrete values and the representation of 
the $c$'s would be finite dimensional. 

In the following we will be interested in a subspace of the full space of states which have the form of being 
 the direct product of the bulk field vacuum $|0>$ and a state  for the operators  operators $c_\sigma(k_2),c_\sigma^\dagger(k_2)$. For now, we will assume that such states exist and that they can be expanded in basis of normalized and orthogonal complex vectors whose elements we will label by a symbol $\gamma$. 
 A basis state in this subspace then has the form 
$$
|\gamma> \equiv |0>\otimes{\rm ~state~of~}c's{\rm ~labeled~by~}\gamma
$$
where  
$$
<\gamma|\gamma'>=\delta_{\gamma,\gamma'}
$$
We will call the basis vectors in this space  $\gamma$-states and the span of the $\gamma$-states the $\gamma$-space.

In all such states, the expectation value of the tight-binding Hamiltonian (\ref{ham1}) is given by the vacuum energy,
\begin{align}
<\gamma|H_0|\gamma'>=E_0\delta_{\gamma,\gamma'}
\label{<H>}
\end{align}
where $E_0$ is given in equation (\ref{E0}). The $\gamma$-space is the space of degenerate eigenstates of $H_0$ with eigenvalue $E_0$. 
They are the degenerate multi-electron
states resulting from populating half of the flat band of edge states. In the next section we will discuss how this degeneracy might be
resolved by interactions amongst the electrons.

\section{Interactions and the resolution of degeneracy}

 We will attempt to resolve the degeneracy of the $\gamma$-states by adding an interaction Hamiltonian to the tight binding Hamiltonian
so that the total Hamiltonian is
\begin{align}\label{ham+hint}
H=H_0+H_{\rm int}
\end{align}
where the interaction Hamiltonian has the form given in equation (\ref{hint}) (which we recopy here for the reader's convenience),
 \begin{align}
 H_{\rm int}=  
\frac{1}{2}\sum_{XY}V(X,Y) \rho(X)\rho(Y)\nonumber  
\end{align}
We will then study the splitting of the degenerate states $|\gamma>$ at the first order in degenerate Rayleigh-Schr\"odinger perturbation theory.    In the first order of this perturbation theory, the corrected  energies of the gamma-states are
given by the eigenvalues of the matrix
\begin{align}\label{curlyH}
\mathcal H_{\gamma\gamma'}~=~E_0\delta_{\gamma,\gamma'}+<\gamma|H_{\rm int}|\gamma'>
\end{align}

The charge density operator which occurs in the interaction Hamiltonian is
\begin{align}
&\rho(X)=\sum_{\sigma=1}^2\psi_\sigma^\dagger(X)\psi_\sigma(X)-1 \nonumber 
\\
&= \sum_{\sigma}\biggl\{ \psi_\sigma^{(0)\dagger}(X)\psi^{(0)}_\sigma(X)  +\psi_\sigma^{(0)\dagger}(X)\tilde\psi_\sigma(X)+\tilde\psi_\sigma^{\dagger}(X)\psi^{(0)}_\sigma(X)
\nonumber \\ &+ \tilde\psi_\sigma^{\dagger}(X)\tilde\psi_\sigma(X)\biggr\}-1  
\end{align}
where we have separated the field operator $\psi_\sigma(X)$ into edge and bulk parts  according to the decomposition in equation (\ref{decomposition}). 
We also recall that the edge part of the field, $\psi_\sigma^{(0)}(X)$, has support only on the $A$ sub-lattice. 
With this expression, the matrix element of the interaction Hamiltonian naturally separates into three parts, 
 \begin{align}
& <\gamma| H_{\rm int} |\gamma'> ~=~ <\gamma| H_{\rm int} |\gamma'>_{\rm edge-edge}\nonumber \\ &
+ <\gamma| H_{\rm int} |\gamma'>_{\rm edge-bulk}+ <\gamma| H_{\rm int} |\gamma'>_{\rm bulk-bulk}
\label{sum}\end{align}
where the edge-edge interaction is 
\begin{align}
& <\gamma| H_{\rm int} |\gamma'>_{\rm edge-edge} ~=~\frac{1}{2}\sum_{AA'}V(A,A')\sum_{\sigma,\rho}\times \nonumber \\ &\times
\biggl<~\gamma~\biggr| \psi_\sigma^{(0)\dagger}(A)\psi^{(0)}_\sigma(A)\psi_\rho^{(0)\dagger}(A')\psi^{(0)}_\rho(A') 
 \biggr|\gamma'\biggr>
  \label{ee}\end{align}
  the edge-bulk interaction is 
  \begin{align}
& <\gamma| H_{\rm int} |\gamma'>_{\rm edge-bulk} ~=~\nonumber \\ 
&\biggl<~\gamma~\biggr| \sum_{A,X}V(A,X)\psi_\sigma^{(0)\dagger}(A)\psi^{(0)}_\sigma(A)[\tilde\psi_\rho^{\dagger}(X)\tilde\psi_\rho(X)-1] 
  \nonumber \\ 
& +\frac{1}{2}\sum_{AA'}V(A,A')\biggl[ \psi_\sigma^{(0)\dagger}(A) \tilde\psi_\sigma(A)\tilde\psi_\rho^{\dagger}(A')\psi^{(0)}_\rho(A')\nonumber \\ &
~~~~~~~~~~~~~~~~~~~~~~+\tilde\psi_\sigma^{\dagger}(A)\psi^{(0)}_\sigma(A)\psi_\rho^{(0)\dagger}(A') \tilde\psi_\rho(A')\biggr] \biggr|\gamma'\biggr>
\label{eb}
\end{align}
where we have dropped some terms which are linear and cubic in $\tilde\psi_\sigma^\dagger(X),~\tilde \psi_\sigma(X)$ because their matrix elements in gamma states vanish. Also, repeated $\sigma$ and $\rho$ indices are assumed to be summed over.

The bulk-bulk interaction is
\begin{align}
& <\gamma| H_{\rm int} |\gamma'>_{\rm bulk-bulk} ~=~E_0\delta_{\gamma,\gamma'}+\nonumber \\ &
\biggl<~\gamma~\biggr|\frac{1}{2}\sum_{XY}V(X,Y)[\tilde\psi_\sigma^{\dagger}(X)\tilde\psi_\sigma(X)-1][ \tilde\psi_\rho^{\dagger}(Y)\tilde\psi_\rho(Y)-1]~ \biggl|~\gamma'~\biggr>
\label{bb}
\end{align}
 
Our next step is to study and try to simplify the three types of matrix elements of the interaction Hamiltonian.  We begin with the edge-bulk interactions in equation (\ref{eb}). In the second line of  (\ref{eb}), $\tilde\psi^\dagger$ and $\tilde\psi$ appear quadratically and they create and re-annihilate a hole. This process does not depend on the label $\gamma$. 
When $X$ is on the $B$ sub-lattice, inside the bracket in equation (\ref{eb}), we can therefore make the replacement
\begin{align}
&\tilde\psi_\sigma^{\dagger}(B)\tilde\psi_\sigma(B)-1\to
2\int_{\Omega^+}dk_1dk_2 \phi^{(-)}(k;B)\phi^{(-)*}(k;B)-1 \nonumber \\ &=
\int_{\Omega^+}dk_1dk_2 [ \phi^{(-)}(k;B)\phi^{(-)*}(k;B) +  \phi^{(+)}(k;B)\phi^{(+)*}(k;B)]\nonumber \\ &~~~~~~~~~-1
\nonumber \\ & =0
\label{id1}\end{align}
where the factor of $2$ in the first line comes from the sum over spin states, we have made use of particle-hole symmetry --  the only difference between a negative and a positive frequency wave-function is a sign-flip of the wave-function on the $A$ sub-lattice -- they are identical on the $B$ sub-lattice and we have used the completeness of the wave-functions, equation (\ref{c1}). 

Alternatively, by similar reasoning, when $X$ is on the $A$ sub-lattice we have 
\begin{align}
&\tilde\psi_\sigma^{\dagger}(A)\tilde\psi_\sigma(A)-1\to
2\int_{\Omega^+}dk_1dk_2  \phi^{(-)}(k;A)\phi^{(-)*}(k;A)-1 \nonumber \\ &=
\int_{\Omega^+}dk_1dk_2 [ \phi^{(-)}(k;A)\phi^{(-)*}(k;A) +  \phi^{(+)}(k;A)\phi^{(+)*}(k;A)]\nonumber \\ &~~~~~~~~-1
\nonumber \\ &= \int dk_2 \phi^{(0)}(k_2;A)\phi^{(0)*}(k_2;A) 
\label{id2}\end{align}
where we have used equation (\ref{c2}).
The replacements (\ref{id1}) and (\ref{id2}) allow us to simplify the second line in equation (\ref{eb}).

We can also simplify the third and fourth lines  in equation (\ref{eb}),
where $\tilde\psi$ and $\tilde\psi^\dagger$ also appear quadratically,  by making the replacements
\begin{align}
&\tilde\psi_\sigma(A)\tilde\psi^\dagger_\rho(A')\to\delta_{\sigma\rho}\int_{\Omega^+}dk_1dk_2 \phi^{(+)}(k;A)\phi^{(+)*}(k;A')   \nonumber \\ &
=\frac{\delta_{\sigma\rho}}{2}\int_{\Omega^+}dk_1dk_2 [ \phi^{(+)}(k;A)\phi^{(+)*}(k;A') +\phi^{(-)}(k;A)\phi^{(-)*}(k;A')]  \nonumber \\ &
=\frac{\delta_{\sigma\rho}}{2} \delta(A,A') - \frac{\delta_{\sigma\rho}}{2} \int dk_2\phi^{(0)}(k_2;A)\phi^{(0)*}(k_2;A') \label{id3}   \\ &
\tilde\psi^\dagger_\sigma(A)\tilde\psi_\rho(A')\to \nonumber \\ &
\frac{\delta_{\sigma\rho}}{2}\delta(A,A')   - \frac{\delta_{\sigma\rho}}{2} \int dk_2 \phi^{(0)}(k_2;A') \phi^{(0)*}(k_2;A) 
\label{id4}
\end{align}
The right-hand-sides of equations (\ref{id3}) and (\ref{id4}) are identical after we relabel $A\leftrightarrow A'$. When we plug them back into
 the third and fourth lines  in equation (\ref{eb}) they become
 \begin{align*}
  &\frac{1}{2}\sum_{AA'}V(A,A')\sum_{\sigma,\rho} \biggl[ \psi_\sigma^{(0)\dagger}(A) \tilde\psi_\sigma(A)\tilde\psi_\rho^{\dagger}(A')\psi^{(0)}_\rho(A')\nonumber \\ &
~~~~~~~~~~~~~~~~~~~~~~+\tilde\psi_\sigma^{\dagger}(A)\psi^{(0)}_\sigma(A)\psi_\rho^{(0)\dagger}(A') \tilde\psi_\rho(A')\biggr] 
\\ &
= \frac{1}{2}\sum_{AA'}V(A,A')\sum_{\sigma,\rho}  \left\{ \psi_\sigma^{(0)\dagger}(A),\psi^{(0)}_\rho(A') \right\}\frac{\delta_{\sigma\rho}}{2}
\times
\\ &
 \times
\biggl[  \delta(A,A')   -  \int dk_2\phi^{(0)}(k_2;A)\phi^{(0)*}(k_2;A')\biggr]
\\ &
= \frac{1}{2}\sum_{AA'}V(A,A')\int dk_2\phi^{(0)*}(k_2;A')\phi^{(0)}(k_2;A)
\times
\\ &
 \times
\biggl[  \delta(A,A')   -  \int dk_2\phi^{(0)}(k_2;A)\phi^{(0)*}(k_2;A')\biggr]
\end{align*}
 where we have used the anti-commutation relation for the edge state field
 \begin{align}\label{anticom}
 \left\{ \psi_\sigma^{(0)\dagger}(A),\psi^{(0)}_\rho(A') \right\}= \delta_{\sigma,\rho}~\int dk_2\phi^{(0)*}(k_2;A)\phi^{(0)}(k_2;A')
 \end{align}
 The sum total of the terms taking into account the interaction of edge and bulk states is thus 
 \begin{align}
&  <\gamma| H_{\rm int} |\gamma'>_{\rm edge-bulk} = -  \biggl<~\gamma~\biggr|\sum_{A,A'}V(A,A')\times  \nonumber \\ &  
\times \psi_\sigma^{(0)\dagger}(A)\psi^{(0)}_\sigma(A)~ \int dk_2 \phi^{(0)*}(k_2;A')\phi^{(0)}(k_2;A')  \biggl|~\gamma'~\biggr>
  \nonumber \\ 
&+\delta(\gamma,\gamma')\biggl\{ \frac{1}{2}\sum_{A}V(A,A)\int dk_2 |\phi^{(0)}(k_2;A)|^2
\\ &
 -  \frac{1}{2}\sum_{AA'}V(A,A')\biggl| \int dk_2\phi^{(0)}(k_2;A)\phi^{(0)*}(k_2;A')\biggr|^2 \biggr\}
\label{int22}
\end{align}
Notice that the last terms which are proportional to $\delta(\gamma,\gamma')$ are otherwise independent of $\gamma$ and $\gamma'$.
They are thus proportional to the unit matrix acting on  the $\gamma$-states. The first term, on the other hand, depends on $\gamma$ and $\gamma'$ as it contains the edge-state density operator.
This part of the edge-bulk interaction is nontrivial and it will play an important role shortly.

Finally, the bulk-bulk term is also completely independent of the $\gamma$-labels and it acts on the $\gamma$-states like the unit matrix times a constant,
\begin{align}
& <\gamma| H_{\rm int} |\gamma'>_{\rm bulk-bulk} ~=~\delta(\gamma,\gamma')\biggl\{\frac{1}{2}\sum_{XY}V(X,Y) \times\nonumber \\ &
\times \biggl[2\int_{\Omega^+}dk_1dk_2  \phi^{(+)}(k;X)\phi^{(+)*}(k;X)-1\biggr]\times \nonumber \\ & \times \biggl[2 \int_{\Omega^+}dk_1'dk_2' \phi^{(-)}(k',Y)\phi^{(-)*}(k';Y)-1\biggr]
\nonumber \\ &
+\frac{1}{2}\sum_{XY}V(X,Y)2 \int_{\Omega^+} dk_1dk_2  \phi^{(+)}(k,X) \phi^{(+)*}(k,Y)\times \nonumber \\ & \times
\int_{\Omega^+} dk_1'dk_2'  \phi^{(-)}(k',Y) \phi^{(-)*}(k',X) \biggr\}
\label{bb1}
\end{align}
where the factors of 2 come from sums over spins. This entire contribution is proportional to the unit matrix in $\gamma$ space and it is actually not needed at all if our only task is to identify the lowest energy states there.  However, as with the edge-bulk contribution, we can simplify terms by using particle-hole symmetry and the completeness relations.
 
Finally, we can recombine the edge-edge, edge-bulk and bulk-bulk contributions to the Hamiltonian to write write the matrix which must be diagonalized to resolve the degeneracy as 
\begin{align}
\mathcal H_{\gamma\gamma'} =& 
\biggl<\gamma\biggr|\frac{1}{2}\sum_{AA'}V(A,A')\rho^{(0)}(A)\rho^{(0)}(A')\biggr| \gamma' \biggr>
 \nonumber \\ &
 +\delta(\gamma,\gamma')(E_0+E^{(1)} )
\label{int3}
\end{align}
where the edge charge density operator is  
\begin{align}
\rho^{(0)}(A)=\sum_{\sigma=\uparrow,\downarrow} \psi^{(0)\dagger}_\sigma(A)\psi_\sigma^{(0)}(A)-\int dk_2| \phi^{(0)}(k_2;A)|^2
\label{density}
\end{align}
where $E_0$ is given in equation (\ref{E0}) and the first correction to the bulk energy is
\begin{align}
 E^{(1)}=&-2\sum_{AB}V(A,B)\biggl| \int_{\Omega^+} dk_1dk_2  \phi^{(+)}(k,A) \phi^{(+)*}(k,B)\biggr|^2
 \nonumber \\ &
-\frac{1}{4}\sum_{AA'}V(A,A')\biggl|\int dk_2 \phi^{(0)}(k_2,A) \phi^{(0)*}(k_2,A')\biggr|^2\nonumber \\ &
+\frac{1}{4}\sum_X V(X,X)
\label{E1}
\end{align}
which, like $E_0$,  does not depend on $\gamma$ and $\gamma'$. We note here that the interaction term for the edge and bulk states was
essential in writing the off-set contribution in the edge charge density operator 
in equation (\ref{density}). 

The task of understanding how the interaction resolves the degeneracy has now been reduced to understanding the eigenvalue spectrum of the
Hermitian matrix $\mathcal H_{\gamma\gamma'}$ in equation (\ref{int3}). This is still a difficult problem which we are
not able to solve in general.  However, we note that the matrix that appears in the first  line of  (\ref{int3}), as well as being Hermitian,  is positive semi-definite.
If we can find a $\gamma$-state which is annihilated by the interaction Hamiltonian, we know that state is an eigenvector of the matrix $\mathcal H$ with the smallest possible eigenvalue.

First, we note that,  since the operator 
$$
h_{\rm int}\equiv \frac{1}{2}\sum_{AA'}V(A,A')\rho^{(0)}(A)\rho^{(0)}(A')
$$
contains only $\psi_\sigma^{(0)}(A)$ and $\psi_\sigma^{(0)\dagger}(A)$, when it
acts on a $\gamma$-state the result is another $\gamma$-state.  
Moreover, since the interaction potential $\mathcal V(X,Y)$ is positive, the smallest possible eigenvalue of the 
above operator is zero.  

To see this, assume that $|F>$ is a normalized eigenvector of $h_{\rm int}$ with eigenvalue $f$.  Then we note that
\begin{align*}
&f=<F|h_{\rm int}|F>=<F|\frac{1}{2}\sum_{AA'}V(A,A')\rho^{(0)}(A)\rho^{(0)}(A')|F> \\
&=\sum_{\gamma'}\frac{1}{2}\sum_{AA'}V(A,A')|<F|\rho^{(0)}(A)|\gamma'><\gamma'|\rho^{(0)}(A')|F>\\
&=\sum_{\gamma',v}\frac{1}{2}v \biggl|  \sum_A \phi_v(A)<\gamma'|\rho^{(0)}(A)|F>\biggr|^2\geq 0
\end{align*}
where we have used equation (\ref{pot2}) for the decomposition of the potential and we recall that positivity of the potential means that $v$ are non-negative
real numbers. Thus $f$ can be zero only when $ \sum_A \phi_v(A)<\gamma'|\rho^{(0)}(A)|F>=0$ for all $v$ and $|\gamma'>$.  Since the set of functions $\{\phi_v(A)\}$ must itself
be complete, we see that $ f$ can be zero only of $<\gamma|\rho^{(0)}(A)|F>=0$ for all $|\gamma>$.  Since $  \rho^{(0)}(A)$ contains only edge state creation and annihilation operators, $\rho^{(0)}(A)|F>$ is a vector in the $\gamma$ space and it must therefore be the zero vector.

Thus, a necessary and sufficient condition for a state to be an eigenstate of $h_{\rm int}$ with zero eigenvalue is 
$$
\rho^{(0)}(A)|F>=0~,~\forall A 
$$
We will   find and characterize states that obey this condition.  For that purpose it is better to realize that $\rho^{(0)}(A)$ are not independent operators of all values of $A$ and to express the condition in terms of its Fourier transform
$$
\sum_{m=1}^Le^{i\frac{2\pi}{\sqrt{3}L}qA_2(m,n)}\rho^{(0)}(A(mn))|F>=0 ~,~A_2(m,n)=\sqrt{3}m-\frac{\sqrt{3}}{2}n
$$
with $q\sim q+L$, Here, we have assumed the periodic identification  (\ref{identification}) and a complete set of plane waves $e^{ik_2A_2}$. 
As a result
\begin{align} 
& \sum_\ell \sqrt{(1-4\cos^2\frac{\pi\ell}{L})(1-4cos^2\frac{\pi(\ell+q)}{L})}\times \nonumber \\ & \times(4\cos\frac{\pi\ell}{L}\cos\frac{\pi(\ell+q)}{L})^n(c_\sigma^\dagger(\ell)c_\sigma(\ell+q)-N\delta_{q0})|F>=0  \label{constraint001} 
\end{align}
where $n$ is a non-negative integer, we recall that the integer $\ell\in (\frac{L}{3},\frac{2L}{3})$, excluding the endpoints and the midpoint,  and the integer $q$ must be such that $\ell+q$ is still an allowed value of the $\ell$'s.   Now, we observe that the coefficient
$$
\sqrt{(1-4\cos^2\frac{\pi\ell}{L})(1-4cos^2\frac{\pi(\ell+q)}{L})} (4\cos\frac{\pi\ell}{L}\cos\frac{\pi(\ell+q)}{L})^n$$
does not uniquely define a pair $\ell,q$  as it is left unchanged by the replacement $\ell\to L-\ell-q$.This allows us to rewrite equation (\ref{constraint001})  as
\begin{align} 
& \sum_\ell \sqrt{(1-4\cos^2\frac{\pi\ell}{L})(1-4cos^2\frac{\pi(\ell+q)}{L})}(4\cos\frac{\pi\ell}{L}\cos\frac{\pi(\ell+q)}{L})^n
\times \nonumber \\ & \times
(c_\sigma^\dagger(\ell)c_\sigma(\ell+q)+c_\sigma^\dagger(L-\ell-q)c_\sigma(L-\ell)-N\delta_{q0})|F>=0  \label{constraint002} 
\end{align} 
Now, we can use the following trick.  We multiply equation (\ref{constraint002}) by $z^{-1-n}$ and sum over $n$.  The we can recover any term in the sum using
the complex line integral 
 \begin{align} 
&\oint_C\frac{dz}{2\pi i} \sum_\ell\frac{ \sqrt{(1-4\cos^2\frac{\pi\ell}{L})(1-4cos^2\frac{\pi(\ell+q)}{L})} }{ z-4\cos\frac{\pi\ell}{L}\cos\frac{\pi(\ell+q)}{L}}
\times \nonumber \\ & \times
(c_\sigma^\dagger(\ell)c_\sigma(\ell+q)+c_\sigma^\dagger(L-\ell-q)c_\sigma(L-\ell)-N\delta_{q0})|F>=0  \label{constraint003} 
\end{align}
where the contour is a small circle enclosing any one of the quantities $4\cos\frac{\pi\ell}{L}\cos\frac{\pi(\ell+q)}{L}$. With use of Cauchy's theorem, this allows us to strip off the summation and
present equation (\ref{constraint002}) as
\begin{align} 
& 
[c_\sigma^\dagger(\ell)c_\sigma(\ell')+c_\sigma^\dagger(L-\ell')c_\sigma(L-\ell)]~|F>=0  ~~(\ell\neq \ell')\label{constraint004} \\ &
[c_\sigma^\dagger(\ell)c_\sigma(L-\ell')+c_\sigma^\dagger( \ell')c_\sigma(L-\ell)]~|F>=0  \label{constraint005} \\ &
 [c_\sigma^\dagger(\ell)c_\sigma(\ell)+c_\sigma^\dagger(L-\ell)c_\sigma(L-\ell)-2]~|F>=0 \label{constraint006} \\ &
\ell,\ell'\in  (\frac{L}{2},\frac{2L}{3})~,~~N=2\sum_\ell 1\nonumber 
\end{align}
where we have separated the $q=0$ constraint in equation (\ref{constraint006}).

Now, we observe that if $\mathcal C_1|F>=0$ and $\mathcal C_2|F>=0$ then $[\mathcal C_1, \mathcal C_2]|F>=0$. Commutators of constraints on $|F>$ are again constraints. Consider a commutator of a constraint of the type in equation (\ref{constraint004}) and of the type in (\ref{constraint006}),
{\small{$$
\left[ 
c_\sigma^\dagger(\ell)c_\sigma(\ell')+c_\sigma^\dagger(L-\ell')c_\sigma(L-\ell) , 
c_\rho^\dagger(\ell)c_\rho(\ell)+c_\rho^\dagger(L-\ell)c_\rho(L-\ell) 
\right]
$$
$$
=-c_\sigma^\dagger(\ell)c_\sigma(\ell') + c_\sigma^\dagger(L-\ell')c_\sigma(L-\ell)
$$}}
which, together with (\ref{constraint004}) implies
\begin{align} 
&
c_\sigma^\dagger(\ell)c_\sigma(\ell')|F>=0,~ c_\sigma^\dagger(L-\ell')c_\sigma(L-\ell)|F>=0
\label{constraint007}\end{align}
for all $\ell\neq\ell'$ and $\ell,\ell'\in (\frac{L}{2},\frac{2L}{3})$. 
A similar argument using a commutator of the constraints in equations (\ref{constraint005}) and (\ref{constraint006}) yields
\begin{align} 
&
c_\sigma^\dagger(\ell)c_\sigma(L-\ell')|F>=0,~ c_\sigma^\dagger(L-\ell)c_\sigma(\ell')|F>=0
\label{constraint008}
\end{align}
The constraints in equations (\ref{constraint007}) and (\ref{constraint008}) coincide with the nonzero roots of an $su(N)$ Lie algebra acting on the $\gamma$-space, where $N$ is two times the number values of $\ell$ in the half-interval.  The Cartan sub-algebra can be gotten by taking commutators of the form 
\begin{align}
&[c_\sigma^\dagger(\ell)c_\sigma(\ell'),c_\rho^\dagger(\ell')c_\rho(\ell)]=
c_\sigma^\dagger(\ell)c_\sigma(\ell')-c_\sigma^\dagger(\ell)'c_\sigma(\ell')
\end{align}
which, when $\ell\neq\ell'$ generates the traceless, diagonal generators of $su(N)$. 

If we revert to the state labeling where $\ell$ and $\ell'$ are in the full interval
$(\frac{L}{3},\frac{2L}{3})$ (excluding the midpoint, if there is one) the set of constraints is 
\begin{align}
&  c_\sigma^\dagger(\ell)c_\sigma(\ell')|F>=0 ~,~\ell\neq \ell' \label{constraint0'}
\\ & [c^\dagger_\sigma(\ell)c_\sigma(\ell)-c^\dagger_\sigma(\ell+1)c_\sigma(\ell+1)] |F>=0 
\label{constraint1'}
 \end{align}
 The operators in equation (\ref{constraint0'}) correspond to all of the non-zero roots and the operators in (\ref{constraint1'}) span the Cartan sub-algebra of the $su(N)$ Lie algebra. 
 The constraints in (\ref{constraint0'}) and (\ref{constraint1'}) require that 
 the state $|F>$   be a trivial, singlet representation of this $su(N)$ Lie algebra.

 The conditions (\ref{constraint006}) in combination with the Cartan sub-algebra elements of $su(N)$ imply that the states are half-filled,
 \begin{align}
[\sum_{\ell} c_\sigma^\dagger(\ell)c_\sigma(\ell)-N]~|F>=0
 \end{align}
 that is, that there are $N$ electrons. 
  
 Let us examine the structure of a state which satisfies these constraints in more detail.   If $|0>$ is the empty band, satisfying
 $$
 c(n)|0>=0~,~\forall n
 $$
 and
 $<0|0>=1$, a half-filled state has $N$ electrons and it must therefore have the generic  form 
 $$
 \sum_{\sigma_1,\ldots,\sigma_N}
 \sum_{\ell_1,\ldots,\ell_N}
 \psi_{\sigma_1\ldots\sigma_N}(\ell_1,\ldots,\ell_N)c^\dagger_{\sigma_1}(\ell_1) \ldots c^{\dagger}_{\sigma_N}(\ell_N)|0>
 $$
 Because the $c^\dagger$'s anti-commute, the coefficients in the superposition,  $ \psi_{\sigma_1\ldots\sigma_N}(\ell_1,\ldots,\ell_N)$, must be completely antisymmetric under simultaneous permutations of the pairs
 of labels $((\sigma_1,\ell_1), (\sigma_2,\ell_2),\ldots, (\sigma_N,\ell_N))$. 
 To be an $su(N)$ singlet, they must be completely antisymmetric under permutations of  $(\ell_1,\ell_2,\ldots,\ell_N)$. This corresponds to the single column Young Tableau with $N$ boxes which is an $su(N)$ singlet.  
 To meet these two criteria,  $ \psi_{\sigma_1\ldots\sigma_N}(\ell_1,\ldots,\ell_N))$  must then be  completely symmetric   in the spin indices $\sigma_1,\sigma_2,\ldots,\sigma_N$.  This means that it transforms under the irreducible $j=N/2$ representation of the $su(2)$ spin algebra. The $2j+1=N+1$ states in this representation 
 comprise   the entire degeneracy of the ground state.
 
 This is the largest irreducible representation of the spin algebra that a collection of $N$ electrons can take up.  It is sometimes called a ``super-spin''.  The highest weight state of the representation is completely polarized
$$
c^\dagger_{1}(\ell_1) \ldots c^{\dagger}_{1}(\ell_N)|0>
$$
and, 
since the representation is irreducible, all of the other states can be obtained by an $su(2)$ rotation of this state.  Therefore all of the possible ground states are polarized in some direction and, in any such state
$$
\left< \sum_A\psi^\dagger(A)\frac{\vec\sigma}{2}\psi(A)\right> = \hat e \frac{N}{2}
$$
where $\hat e$ is a unit vector in the direction of the spin polarization.
The set of ground states are a three-dimensional quantum rotor which becomes classical in the limit $N\to\infty$.  This is the sense in which it is a ferromagnet.

\section{Discussion}
\label{discussion}

We have shown that, for a  large class of weak repulsive two-body interactions, the degeneracy of the flat band of edge states of  zigzag edged graphene is resolved by the interaction in such a way that the lowest energy state is a ferromagnet.  The remaining degeneracy of this ground state is simply the degeneracy of a spin $j=\frac{N}{2}$ representation of the $su(2)$ spin algebra. 

This demonstration contained an interesting relationship between the lowest energy states of the Hamiltonian, which in this case are states with vanishing charge density, and a certain unitary symmetry which is a subgroup of the set of all unitary transformations acting on the multi-electron states in partial fillings of the flat band of edge states. 

 It would be interesting to apply our completeness technique to  excited states in the bulk, as they are also highly degenerate. 
 
\appendix

 \section{ Lattice and dual lattice: notation and conventions}
 \label{lattice}
  
 Let us briefly review some details about the hexagonal graphene lattice.  The   lattice  is depicted in figure \ref{fig}. 
 It is a hexagonal lattice and it has an edges on the left-hand-side of the figure.  The edge is of zig-zag type.  The lattice is assumed to be semi-infinite, to continue indefinitely to the right.  In the up and down directions the lattice it could also be infinite, although  we will also take it as having  a periodic identification.  
 
 The hexagonal lattice is  a superposition of two 
 triangular sub-lattices which we call the $A$ and $B$ sub-lattices.  Any point of the $A$ sub-lattice has three nearest neighbours which are on the $B$
sub-lattice and vice versa. The $A$ sub-lattice sites are the blue dots and the $B$ sub-lattice sites are the red dots in figure \ref{fig}.  
In a system of distance units where the lattice constant is equal to one, the two sub-lattices are connected by the three unit vectors
 \begin{align}
 &\hat \delta_1=(-1,0),~\hat \delta_2 = (\frac{1}{2},\frac{\sqrt{3}}{2}),~\hat \delta_3=  (\frac{1}{2},-\frac{\sqrt{3}}{2})\\
&\hat \delta_i^2=1,~\hat \delta_i\cdot  \hat \delta_j =-\frac{1}{2},~i\neq j 
 \end{align}
 which obey $\hat \delta_1+\hat \delta_2+\hat \delta_3=0$. The vectors $\hat\delta_i$  originate on $A$ sites and end on neighbouring $B$ sites.
 
 Either of the  $A$ or $B$ sub-lattices is generated by  any two of the following three  vectors
 \begin{align}
 &\vec a_1=\hat \delta_2-\hat \delta_3= (0 ,\sqrt{3}) \\
& \vec a_2= \hat \delta_3-\hat \delta_1=(\frac{3}{2}, \frac{-\sqrt{3}}{2})\\
 &\hat a_3= \hat \delta_1-\hat \delta_2=(-\frac{3}{2},-\frac{\sqrt{3}}{2})
 \end{align}
 We can thus take the sites of the $A$ and $B$ sub-lattices as points on the two dimensional plane with Cartesian coordinates 
 given by
 \begin{align}
&A=m\vec a_1+n\vec a_2-\hat \delta_1
=\biggl(\frac{3}{2}n+1 , \sqrt{3}m-\frac{\sqrt{3}}{2}n \biggr)\label{a}
\\
&B=m\vec a_1+n\vec a_2
=\biggl(\frac{3}{2}n,\sqrt{3}m-\frac{\sqrt{3}}{2}n \biggr)\label{B}
 \end{align}
For convenience, we have chosen the origin of the coordinate system so that the lattice is symmetric under putting $B_1\to -B_1$, implemented by $n\to-n$.  This will be convenient for imposing a  boundary condition  for the zigzag edge where the wave-function must vanish when $B_1=0$. 
 
 When the lattice is the entire two-dimensional plane, $m$ and $n$ in equations (\ref{a}) and (\ref{B}) run over the integers.  
 When it is a right-hand ($x>0$) half of the xy-plane with a zigzag edge, $m$ runs over the integers and $n=0,1,2,\ldots$. When it has a periodic identification in the $A_2,B_2$ directions and the zigzag edge, $m=0,1,2,\ldots,L-1$ and $n=0,1,2,\ldots$. 
  
    The zig-zag edge, on the left-hand-side in figure \ref{fig}, has sites are located on the $A$ sub-lattice at  $n=0$, that is, at $A=(1, \sqrt{3}m )$.  
 The boundary condition for the tight-binding model  is such that the wave-function must 
vanish on the $B$ sites at $n=0$, that is where $B=\left(0,\sqrt{3}m  \right)$.

 The dual of   the $A$   sub-lattice is generated by vectors $b_i$ which obey the equation
  \begin{align}
  e^{i b_i \cdot   A}=1,~\forall A
\end{align}
The generators are easily found to be
\begin{align}
 & b_1= -\frac{4\pi}{3} \hat \delta_1,~
 b_2=-\frac{4\pi}{3} \hat \delta_2,~
 b_3 = - \frac{4\pi}{3} \hat \delta_3 
  \label{b}\end{align}
  These also generate the dual of the $B$ sub-lattice. 

The Brillouin zone is a unit cell of the dual lattice which is usually taken as a hexagon centred on the origin, $\vec k=0$, and having vertices on the $K$-points which are
the solutions of the equation
\begin{align}
\sum_i e^{i\vec K\cdot  \hat \delta_i}=0 
\label{K}
\end{align}
This equation is equivalent to $S(k)=0$ where $S(k)$, defined in equation (\ref{s}),  is the Fourier representation of the displacement operator which appears
in the tight-binding Hamiltonian (\ref{ham}).  These are the points where the negative and positive frequency bands intersect and in charge neutral graphene on the infinite plane, they are also at the Fermi level.  It is a linearization of the frequency spectrum about these points which gives graphene its low energy Dirac fermions. 

The solutions of equation (\ref{K}) occur when the three complex numbers $e^{i\vec K\cdot  \hat \delta_i}$ are the three cube roots of unity, so that, for example,
\begin{align}
\vec K\cdot   \hat \delta_1=0,~\vec K\cdot  \hat \delta_2=\frac{2\pi}{3},~\vec K\cdot  \hat \delta_3=-\frac{2\pi}{3}
\end{align}
which is solved by
\begin{align}
\vec K=\frac{4\pi}{9}(\hat \delta_2-\hat \delta_3)
\end{align}
The complete list of such $K$-points is
\begin{align}
&\vec K_1= \frac{4\pi}{9}(\hat \delta_2-\hat \delta_3),~
 \frac{4\pi}{9}(\hat \delta_3-\hat \delta_1),~
 \frac{4\pi}{9}(\hat \delta_1-\hat \delta_2) \label{k1}\\
&\vec K_2=- \frac{4\pi}{9}(\hat \delta_2-\hat \delta_3),~
-\frac{4\pi}{9}(\hat \delta_3-\hat \delta_1),~
-\frac{4\pi}{9}(\hat \delta_1-\hat \delta_2)  \label{k2}
\end{align}
It is easy to see that, since  the projection of a $K$-point along the direction of a nearby dual lattice basis vector, for example
$$
\vec K_1\cdot   b_3/|b_3|= \frac{4\pi}{9}(\hat \delta_2-\hat \delta_3)(- \hat\delta_3 )=2\pi/3
 $$
 is half of the length $|b_3|=4\pi/3$ of the dual lattice basis vector, the hexagon with vertices on the $K$-points is indeed the Wigner-Seitz cell of
 the dual lattice, which is the usual choice for the first Brillouin zone.
 
We can see that a difference of any two of the $K_1$-points is a sum of dual lattice vectors,
for example,
\begin{align}&
 \frac{4\pi}{9}(\vec \delta_2-\vec \delta_3)-\frac{4\pi}{9}(\vec \delta_3-\vec \delta_1)
 =\frac{4\pi}{9}(-2\vec \delta_3+\vec \delta_1+\vec \delta_2)
 \nonumber \\ &
 =\frac{4\pi}{9}(-3\vec \delta_3)=b_3
 \end{align}
 This is also the case for the $K_2$-points. This means that there are only two independent $K$-points, for which we 
could choose any one from the $K_1$ list (\ref{k1}) and any one from the $K_2$ list (\ref{k2}).

The lattice Fourier transform on the infinite plane depends on the completeness and orthogonality of plane waves, 
\begin{align}
\sum_A e^{i\vec k\cdot   A} 
=\Omega\delta(k,0)
,~
\int_{\Omega}d^2k  e^{i\vec k\cdot  A}=\Omega \delta(A,0)\label{completeness1}\\
\sum_B e^{i\vec k\cdot   B} 
=\Omega\delta(k)
,~
\int_{\Omega}  d^2k e^{i\vec k\cdot  B}=\Omega  \delta(B,0) 
\label{completeness2}
\end{align}
where we use the symbol $\Omega$ for both the Brillouin zone and its volume which
 appears as a factor on the right-hand-sides of these equations.

\section{Completeness of wavefunctions}
\label{completeness}

In the following we will demonstrate completeness of the wave-functions with the edge states included. We will consider the case on the $A$ sub-lattice
which is the most complicated one.  
Consider the completeness integral
\begin{align}
&\Delta(A,A')\equiv \nonumber \\ 
&\int_{\Omega^+}dk_1dk_2\biggl[  \phi^{(+)}(k;A)\phi^{(+)*}(k,A')+   \phi^{(-)}(k;A)\phi^{(-)*}(k,A')\biggr]
\nonumber \\ &
=2\int_{\Omega^+}dk_1dk_2
\frac{ e^{ik_2A_2} }{\sqrt{2\Omega} }\biggl[ e^{ik_1A_1}\frac{tS(k)}{|S(k)|} - e^{-ik_1A_1}\frac{tS^*(k)}{|S(k)|}  \biggr]\times
\nonumber \\ &
\times
\frac{ e^{-ik_2A_2'} }{\sqrt{2\Omega} }\biggl[ e^{-ik_1A_1'}\frac{t^*S^*(k)}{|S(k)|} - e^{ik_1A_1'}\frac{t^*S(k)}{|S(k)|}  \biggr]
\end{align}
By combining terms, we can restore the integral over the full Brillouin zone, 
\begin{align}
&\Delta(A,A')\equiv \nonumber \\ 
&=\frac{1}{\Omega}\int_{\Omega}dk_1dk_2
e^{ik_2(A_2-A_2')} \biggl[ e^{ik_1(A_1-A_1')} - e^{ik_1(A_1+A_1')} \frac{S(k)}{S^*(k)}   \biggr] 
\nonumber \\ &
=\delta(A,A') -\nonumber \\ & \frac{1}{\Omega}\int_{\Omega}dk_1dk_2e^{ik_2(A_2-A_2')}
  e^{i \frac{3}{2}k_1(n+n')+ 2ik_1 } \frac{ e^{-ik_1}+e^{ik_1/2}2\cos \frac{\sqrt{3}}{2}k_2 }{ e^{ik_1}+e^{-ik_1/2}2\cos\frac{\sqrt{3}}{2}k_2 }  
\label{comp}\end{align}
We have obtained the $A$ sub-lattice delta function that we expected minus a deficit term.  By defining the complex variable $z= e^{i\frac{3}{2}k_1/2}$ and noting
that the range of integration over $k_1$ is precisely such that $z$ wraps the unit circle once with counter-clockwise orientation, we can write this deficit term as a
contour integral around the unit circle,
\begin{align}
& - \frac{1}{\Omega}\int_{\Omega}dk_1dk_2e^{ik_2(A_2-A_2')}
  e^{i \frac{3}{2}k_1(n+n')  } \frac{ e^{i\frac{3}{2}k_1/2}+e^{3ik_1}2\cos \frac{\sqrt{3}}{2}k_2 }{ e^{i\frac{3}{2}k_1}+ 2\cos\frac{\sqrt{3}}{2}k_2 }  
\nonumber \\ &
= - \frac{1}{\Omega}\frac{2}{3i}\int dk_2e^{ik_2(A_2-A_2')}\oint\frac{dz}{z}
  z^{n+n'}   \frac{ z+z^2 2\cos \frac{\sqrt{3}}{2}k_2 }{ z+ 2\cos\frac{\sqrt{3}}{2}k_2 }  
\label{comp1}\end{align}
where the contour integral is over the unit circle. Using Cauchy's theorem we get
$$
  - \frac{1}{\Omega}\frac{4\pi}{3}\int dk_2 e^{ik_2(A_2-A_2')}
[-2\cos\frac{\sqrt{3}}{2}k_2]^{n+n'}  [1- 4\cos^2 \frac{\sqrt{3}}{2}k_2 ] 
$$
 The integral over $k_2$ in the equation above is over those values  of $k_2$  
where the pole in the contour integral  that was used to obtain this formula is inside the unit circle, that is where
$$
-1<2\cos  \frac{\sqrt{3}}{2}k_2<1
$$
  This is
identical to region of $k_2$ where the edge states are defined.
The 
factor in front is $ \frac{1}{\Omega}\frac{4\pi}{3}$ which, with $\Omega$ from equation (\ref{Omega}) is equal to $\sqrt{3}/2\pi$ which matches the square of the normalization of the 
edge state wave-functions given in equation (\ref{e}).  The integrand is equal to a product of edge states and the integral over $k$ is the integration over all edge states. Then the above expression is identical to $(-1)$ times the completeness sum over the edge states. 
The upshot of equation (\ref{comp}) thus becomes the completeness relation
\begin{align}
&
\int_{\Omega^+}dk_1dk_2\biggl[  \phi^{(+)}(k;A)\phi^{(+)*}(k,A')+   \phi^{(-)}(k;A)\phi^{(-)*}(k,A')\biggr]
\nonumber \\
& +\int dk_2\phi^{(0)}(k;A)\phi^{(0)*}(k_2,A')=\delta(A,A')
 \end{align}

  There is some arbitrariness in the choice of the Brillouin zone.  It must be a fundamental cell of the dual lattice and it is usually taken to be the hexagon with vertices
the $K$-points which we have described above. 
For performing integrals, it is more convenient to use an equivalent Brillouin zone which is the rectangle 
\begin{align}
\Omega=\biggl\{(k_1,k_2)\biggl|-\frac{2\pi}{3}<k_1\leq\frac{2\pi}{3},-\frac{2\pi}{3}\frac{1}{\sqrt{3}}<k_2\leq \frac{4\pi}{3}\frac{1}{\sqrt{3}}\biggr\}
\end{align}
The volume of the Brillouin zone is the area of this rectangle,
\begin{align}\label{Omega}
\Omega=\frac{4\pi}{3}\cdot\frac{2\pi}{\sqrt{3}}
\end{align}
and this is identical to the area of the hexagon.  For the latticized half-plane, we also have taken the range of the wave-umbers to be in the half-zone $\Omega^+$ which could be taken to be
\begin{align}
\Omega^+=\biggl\{(k_1,k_2)\biggl|0\leq k_1<\frac{2\pi}{3},-\frac{2\pi}{3}\frac{1}{\sqrt{3}}<k_2\leq \frac{4\pi}{3}\frac{1}{\sqrt{3}}\biggr\}
\end{align}
The wave-functions for the positive and negative frequency bands
are parameterized by a wave-number $k$ which takes values on $\Omega^+$.

 \end{document}